\documentclass[prl,reprint,amsmath,amssymb,aps,preprintnumbers, longbibliography]{revtex4-1}

\usepackage[colorlinks=true,linkcolor=black, citecolor=black,
urlcolor=black]{hyperref}

\usepackage{multirow,graphics}
\usepackage{enumerate}
\usepackage{amstext}
\usepackage{amssymb}
\usepackage{amsmath}
\usepackage{graphicx}
\usepackage{color}
\usepackage{tikz}
\usepackage{psfrag}
\usepackage{soul}

\usepackage{thumbpdf}
\usepackage{esint,shuffle,psfrag}
 \usepackage[utf8]{inputenc}

\allowdisplaybreaks
\usepackage{varioref}
\usepackage{amsmath,amsfonts,amsthm,mathrsfs}
\usepackage{enumerate}
\usepackage{fancyvrb}
\usepackage{verbatim}
\usepackage{wrapfig}
\usepackage{appendix}
\usepackage{amstext}
\usepackage{amssymb}
\usepackage{graphicx}
\usepackage{color}
\usepackage{varioref}
\usepackage{multirow,graphics}
\usepackage{epstopdf}
\usepackage{bbm}
\usepackage{slashed}
\usepackage{psfrag}
\usepackage{relsize}
\usepackage{bbm}
\usepackage{mathrsfs}
\usepackage{epsfig}
\usepackage{upgreek}

\usepackage{mathtools}
       \usepackage{dsfont}

\setcitestyle{square,numbers}

\begin{document}

\newcommand{\IM}{{\rm Im}\,}
\newcommand{\card}{\#}
\newcommand{\la}[1]{\label{#1}}
\newcommand{\eq}[1]{(\ref{#1}),} 
\newcommand{\figref}[1]{Fig. \ref{#1}}
\newcommand{\abs}[1]{\left|#1\right|}
\newcommand{\comD}[1]{{\color{red}#1\color{black}}}
\newcommand{\p}{{\partial}}
\newcommand{\Tr}{{\text{Tr}}}
\newcommand{\tr}{{\text{tr}}}
\newcommand{\sym}{${\cal N}=4$ SYM becomes a non-unitary and non-supersymmetric CFT. }
\newcommand{\como}[1]{{\color[rgb]{0.0,0.1,0.9} {\bf \"O:} #1} }

\makeatletter
\newcommand{\subalign}[1]{%
  \vcenter{%
    \Let@ \restore@math@cr \default@tag
    \baselineskip\fontdimen10 \scriptfont\tw@
    \advance\baselineskip\fontdimen12 \scriptfont\tw@
    \lineskip\thr@@\fontdimen8 \scriptfont\thr@@
    \lineskiplimit\lineskip
    \ialign{\hfil$\m@th\scriptstyle##$&$\m@th\scriptstyle{}##$\crcr
      #1\crcr
    }%
  }
}
\makeatother

\newcommand{\mzvv}[2]{
  \zeta_{
    \subalign{
      &#1,\\
      &#2
    }
}
}

\newcommand{\mzvvv}[3]{
  \zeta_{
    \subalign{
      &#1,\\
      &#2,\\
      &#3
    }
}
  }

\makeatletter
     \@ifundefined{usebibtex}{\newcommand{\ifbibtexelse}[2]{#2}} {\newcommand{\ifbibtexelse}[2]{#1}}
\makeatother

\preprint{ZMP--HH/19--27}


\usetikzlibrary{decorations.pathmorphing}
\usetikzlibrary{decorations.markings}
\usetikzlibrary{intersections}
\usetikzlibrary{calc}

\tikzset{
photon/.style={decorate, decoration={snake}},
particle/.style={postaction={decorate},
    decoration={markings,mark=at position .5 with {\arrow{>}}}},
antiparticle/.style={postaction={decorate},
    decoration={markings,mark=at position .5 with {\arrow{<}}}},
gluon/.style={decorate, decoration={coil,amplitude=2pt, segment length=4pt},color=purple},
wilson/.style={color=blue, thick},
scalarZ/.style={postaction={decorate},decoration={markings, mark=at position .5 with{\arrow[scale=1]{stealth}}}},
scalarX/.style={postaction={decorate}, dashed, dash pattern = on 4pt off 2pt, dash phase = 2pt, decoration={markings, mark=at position .53 with{\arrow[scale=1]{stealth}}}},
scalarZw/.style={postaction={decorate},decoration={markings, mark=at position .75 with{\arrow[scale=1]{stealth}}}},
scalarXw/.style={postaction={decorate}, dashed, dash pattern = on 4pt off 2pt, dash phase = 2pt, decoration={markings, mark=at position .60 with{\arrow[scale=1]{stealth}}}}
}

 \newcommand{\doublewheelsmall}{
   \begin{minipage}[c]{1cm}
     
     \begin{center}
       \begin{tikzpicture}[scale=0.3]
         \foreach \m in {1,2} {
           \draw (0.75*\m,0) arc[radius = 0.75*\m,start angle = 0, end angle = 300] ;
           \draw[black, densely dotted] (300:0.78*\m) arc[radius = 0.75*\m,start angle = -60, end angle = 0];
         }
         \foreach \t in {1,2,...,5} {
           \draw (0,0) -- (60*\t:1.5);
         }
         \draw[black,densely dotted] (0,0) -- (0:1.5);
       \end{tikzpicture}
     \end{center}
   \end{minipage}
 }


\newcommand{\footnoteab}[2]{\ifbibtexelse{%
\footnotetext{#1}%
\footnotetext{#2}%
\cite{Note1,Note2}%
}{%
\newcommand{\textfootnotea}{#1}%
\newcommand{\textfootnoteab}{#2}%
\cite{thefootnotea,thefootnoteab}}}

\def\e{\epsilon}
     \def\bT{{\bf T}}
    \def\bQ{{\bf Q}}
    \def\wT{{\mathbb{T}}}
    \def\wQ{{\mathbb{Q}}}
    \def\ttQ{{\bar Q}}
    \def\tQ{{\tilde \bP}}
        \def\bP{{\bf P}}
        \def\dq{{\dot q}}
    \def\CF{{\cal F}}
    \def\cC{\CF}
    
     \def\l{\lambda}
\def\hbZ{{\widehat{ Z}}}
\def\bZ{{\resizebox{0.28cm}{0.33cm}{$\hspace{0.03cm}\check {\hspace{-0.03cm}\resizebox{0.14cm}{0.18cm}{$Z$}}$}}}

\title{Exactly solvable magnet of conformal spins in four dimensions}

\author{Sergey Derkachov$^{a}$ and\, Enrico Olivucci$^{b}$}

\affiliation{%
\\
\(^{a}\) St.~Petersburg Department of the Steklov Mathematical Institute, 191023 St.~Petersburg, Russia
 \\
\(^{b}\) 
II. Institute for Theoretical Physics of the University of Hamburg, 22761
Hamburg, Germany
\\
}

\begin{abstract}
We provide the eigenfunctions for a quantum chain of $N$ conformal spins with nearest-neighbor interaction and open boundary conditions in the irreducible representation of $SO(1,5)$ of scaling dimension $\Delta = 2 - i \lambda$ and spin numbers $\ell=\dot{\ell}=0$. The spectrum of the model is separated into $N$ equal contributions, each dependent on a quantum number $Y_a=[\nu_a,n_a]$ which labels a representation of the principal series. The eigenfunctions are orthogonal and we computed the spectral measure by means of a new star-triangle identity. Any portion of a conformal Feynmann diagram with square lattice topology can be represented in terms of separated variables, and we reproduce the all-loop ``fishnet" integrals computed by B.~Basso and L.~Dixon via bootstrap techniques. We conjecture that the proposed  eigenfunctions form a complete set and provide a tool for the direct computation of conformal data in the fishnet limit of the supersymmetric $\mathcal{N}=4\,$  Yang-Mills theory at finite order in the coupling, by means of a cutting-and-gluing procedure on the square lattice.
\end{abstract}  

 \maketitle

\section{Introduction}

The exactly solvable spin magnets \cite{Bethe1931,Baxter1982} constitute a class of condensed matter models of wide interest throughout theoretical and mathematical physics. In particular, the integrable chains of nearest-neighbors interacting spins \cite{Korepin1997,Faddeev2016} serve as a tool to encode the symmetries of local or non-local operators in quantum field theory, providing a rich amount of non-perturbative results ranging from the scattering spectrum of high-energy gluons in QCD  \cite{Fadin1975,Balitskii1978,Lipa:1993pmr} to the conformal data of the super-symmetric $\mathcal{N}=4$ SYM and $\mathcal{N}=6$ ABJM theories~\cite{Beisert:2010jr}.
The archetype model of this class is the $SU(2)$ Heisenberg magnet of spin $\frac{1}{2}$, which for open boundary conditions is described by the Hamiltonian
\begin{align}
\label{XXX}
H_{SU(2)} = \sum_{a=1}^{N-1} \,\vec \sigma_a \cdot \vec \sigma_{a+1}\,,
\end{align} 
being $\vec \sigma_a$ the vector of Pauli matrices acting on the space $\mathbb{V}_a = \mathbb{C}^2$. Generalizations of \eqref{XXX} to other symmetry groups are known, including the non-compact $SO(1,5)$ spin chain \footnote{We consider an euclidean space-time in the letter, without loss of generality respect to the minkowskian case.}. The latter model is relevant for the study of covariant quantities in a four-dimensional conformal field theory (CFT) \cite{DiFrancesco1997}. We consider the homogeneous model in the irreducible unitary representation defined by the scaling dimension $\Delta=2-i\lambda,\,\,\lambda \in \mathbb{R},$ and the $SO(4)$ spins $\ell=\dot{\ell}=0$ \cite{Dobrev1977}. The Hamiltonian operator acts on the Hilbert spaces $\mathbb{V}_a = L^2(x_a,d^4x_a)$ as
\begin{align}
\label{4Dlocal}
\mathbb{H} =&\sum_{a=1}^{N-1} \left[2 \ln x^2_{aa+1} +{(x^2_{aa+1})^{-i\lambda}}\ln(\hat p^2_a\hat p^2_{a+1}) (x^2_{aa+1})^{i\lambda}
 \right]+\notag \\&+ 2\ln x^2_{N0} + \ln(\hat p^2_1)+ (x_{N0}^2)^{-i\lambda}\ln(p_N^2)(x_{N0}^2)^{i\lambda}\,,
\end{align}
where $x_{aa+1}=x_a-x_{a+1}$, $\hat p^2_{a} = -\partial_a \cdot\partial_{a}$ and $x_{N+1}=x_0$. The point $x_0$ is effectively a parameter for the model, and we will always omit it from the set of coordinates. 
The spin chain \eqref{4Dlocal} is the four-dimensional version of the open $SL(2,\mathbb{C})$ Heisenberg magnet which describes the scattering amplitudes of high energy gluons in the Regge limit of QCD \cite{Lipa:1993pmr,Lipatov2004}. The integrability of \eqref{4Dlocal} is realized by the commutative family of normal operators 
\begin{align}
\label{TN}
\mathbb{Q}_N(u)= \text{Q}_{12}(u)\cdot \text{Q}_{23}(u)\cdots \text{Q}_{N0}(u)\, ,
\end{align}
labeled by the spectral parameter $u\in \mathbb{R}$  and where
\begin{equation}
\notag
\text{Q}_{ij}(u)= (x_{ij}^2)^{-i\lambda}(\hat p_i^2)^{u}(x_{ij}^2)^{u+i\lambda}\,.
\end{equation}
By the introduction of the operator
\begin{align*}
\widehat{\mathbb{Q}}_N(u) = [\mathbb{Q}_N(u-i\lambda)]^{\dagger}\mathbb{Q}_N(-i\lambda)\,,
\end{align*}
the Hamiltonian $\mathbb{H}$ is recovered from the expansion
\begin{equation}
\label{Texp}
\mathbb{Q}_N(u)+ \widehat{\mathbb{Q}}_N(u)=2 \cdot \mathbbm{1}+ u\,\mathbb{H}  + o(u)\,.
\end{equation} It follows from \eqref{Texp} and from the commutation relation $[\mathbb{Q}_N(u),\mathbb{Q}_N(v)]=0$ at generic $u$ and $v$, that the eigenfunctions of $\mathbb{Q}_N$ diagonalize the Hamiltonian \eqref{4Dlocal} as well.
The spectra of these operators are labeled by the quantum numbers
\begin{equation}
\label{SOV}
\medmuskip=3mu
\thinmuskip=5mu
\thickmuskip=3mu
Y_a = 1+\frac{n_a}{2} + i \nu_a,\,\,Y^*_a =1+ \frac{n_a}{2} -i \nu_a ,\;\; \nu_a \in \mathbb{R}\,,\; n_a\in \mathbb{N}\,,
\end{equation}
for $a=1,\dots ,N$, and we use to write $\mathbf{Y}=(Y_1,\dots,Y_N)$. The spectral equation for the operator \eqref{TN} reads
\begin{align}
\notag
\mathbb{Q}_N(u) \cdot  \Psi^{\boldsymbol{\alpha\beta}}(\mathbf{x}|\mathbf{Y})= \tau_N(u,\mathbf{Y})\,\Psi^{\boldsymbol{\alpha\beta}}(\mathbf{x}|\mathbf{Y})\,,
\end{align}
where we denote $\mathbf{x}=(x_1,\dots ,x_N)$ and $\boldsymbol{\alpha}$,$\boldsymbol{\beta}$ stand for $2N$ auxiliary complex spinors \begin{equation}
\notag
|\alpha_1\rangle ,\dots ,|\alpha_N\rangle\,\,\,\text{and}\,\,\,|\beta_1\rangle,\dots,\,|\beta_N\rangle\,\in \mathbb{C}^2.
\end{equation} The eigenfunctions form an orthogonal set respect to the quantum numbers $(\mathbf{Y},\boldsymbol{\alpha},\boldsymbol{\beta})$, and the eigenvalue is factorized respect to the labels \eqref{SOV} into equal contributions
\begin{align}
\label{tauN}
\tau_N(u,\mathbf{Y})&= \prod_{a=1}^N \tau_1(u,Y_a)\,,\\
\notag
 \tau_1(u,Y_a)&= 4^u \frac{\Gamma\left(Y_a-\frac{i }{2}\lambda  \right)\Gamma \left(Y_a^*+u+\frac{i }{2}\lambda \right)}{\Gamma \left(Y_a^*+\frac{i}{2}\lambda \right) \Gamma \left(Y_a-u-\frac{i }{2}\lambda \right)}\,.
\end{align}
As a consequence of \eqref{Texp} and \eqref{tauN} we obtained the spectrum of the Hamiltonian $\mathbb{H}$ as a sum of $N$ independent terms
\begin{equation}
\medmuskip=2mu
\thinmuskip=2mu
\thickmuskip=2mu
\label{etaN}
\eta_N(\mathbf{Y})=\sum_{a=1}^N \left[\psi\left(Y_a-\frac{i }{2}\lambda \right)+\psi\left(Y_a+\frac{i }{2}\lambda 
   \right)+ \ln 4 \right] + \,\text{c.c.} \,
\end{equation}
Formulas \eqref{tauN},\eqref{etaN} show that the $N$-body system defined in \eqref{4Dlocal} gets separated into $N$ one-particle systems over the quantum numbers \eqref{SOV}.
In other words, the quantities $(Y_a,|\alpha_a\rangle,|\beta_a\rangle)$ are the separated variables of the system in the sense of \cite{Sklyanin1989,Sklyanin1991a,Sklyanin1995,Sklyanin1996}, and the spectrum of \eqref{4Dlocal} and \eqref{TN} is degenerate in the spinors due to rotation invariance.\\
\\
The representation over the separated variables $(\mathbf{Y},\boldsymbol{\alpha},\boldsymbol{\beta})$ is defined for a generic function $\phi(\mathbf{x})= \phi(x_1,\dots,x_N)$ by the linear transform
\begin{equation}
\label{tilphi}
\widetilde{\phi}(\mathbf{Y},\boldsymbol{\alpha},\boldsymbol{\beta})  =\int d\mathbf{x}\, \Psi^{\boldsymbol{\alpha\beta}}(\mathbf{x}|\mathbf{Y})^* \,\phi(\mathbf{x})\,.
\end{equation}
The inverse transform of \eqref{tilphi} provides the expansion of $\phi(\mathbf{x})$ over the basis of eigenfunctions
\begin{equation}
\label{Ytrans}
\medmuskip=1mu
\thinmuskip=0.7mu
\thickmuskip=0.7mu
\phi(\mathbf{x})= \sum_{\mathbf{n}} \int d\boldsymbol{\nu} \,\,\,\,\,\,\,\,\mu(\mathbf{Y}) \int D\boldsymbol{\alpha}D\boldsymbol{\beta} \,\,\,\,\,\,\,\,\,\,\,\,\,\Psi^{\boldsymbol{\alpha\beta}}(\mathbf{x}|\mathbf{Y})\,\,\,\,\, \widetilde{\phi}(\mathbf{Y},\boldsymbol{\alpha},\boldsymbol{\beta})\,\,\,\,,
\end{equation}
where the sum runs over the non-negative integers $\mathbf{n}=(n_1,\dots,n_N)$, the integrations $d\boldsymbol{\nu}=d\nu_1\cdots d\nu_N$ are on the real line and the integration in the space of spinors $D\boldsymbol{\alpha} = D\alpha_1 \cdots D\alpha_N$ is defined as
\begin{equation}
\notag
\int D \alpha = \int_{\mathbb{C}^2} d\alpha\,e^{-\langle \alpha |\alpha \rangle}\,,\,\,\,\,\,\langle\alpha|\alpha \rangle = |\alpha^{(1)}|^2+|\alpha^{(2)}|^2\,.
\end{equation}
The spectral measure in \eqref{Ytrans} can be extracted from the scalar product of eigenfunctions and it is given by
\begin{equation}
\label{BDmeas}
\medmuskip=0.2mu
\thinmuskip=0.5mu
\thickmuskip=0.2mu
\mu(\mathbf{Y})=\frac{1}{N!} \prod_{a=1}^N{(n_a+1)}\prod_{b\neq a}^N \left[\nu_{ab}^2+\frac{n_{ab}^2}{4}\right]\left[\nu_{ab}^2+\frac{(n_a+n_b+2)^2}{4}\right]\,,
\end{equation}
in the notation $\nu_{ab}=\nu_a-\nu_b$ and $n_{ab}=n_a-n_b$.\\\\
All considerations done so far can be extended by an accurate analytic continuation of the parameter $\lambda$ to the imaginary strip $(-2i,+2i)$. In particular, at $\lambda=-i$ each site of the chain carries the representation $\Delta=1$, $\ell=\dot{\ell}=0$ of a bare scalar field in four dimensions. In this case at the point $u=-1$ the operator $\mathbb{Q}_N(u)$ becomes proportional to the graph-building integral operator for a Feynmann diagram of square lattice topology 
\begin{align}
\label{graphB}
\mathbb{B}_N \,\phi(\mathbf{x}) =\frac{1}{(2\pi)^{4N}}\int d\mathbf{x'} \phi(\mathbf{x'})\prod_{a=1}^N \frac{1}{x_{aa+1}^2 \,x_{aa'} ^2}\,,
\end{align}
with $\mathbf{x}=(x_1,\dots,x_N)$, $\mathbf{x'}=(x'_{1},\dots,x'_{N})$. Throughout the letter we denote $x_{ab'}= x_a-x'_b$.
\begin{figure}
 \includegraphics[scale=0.27]{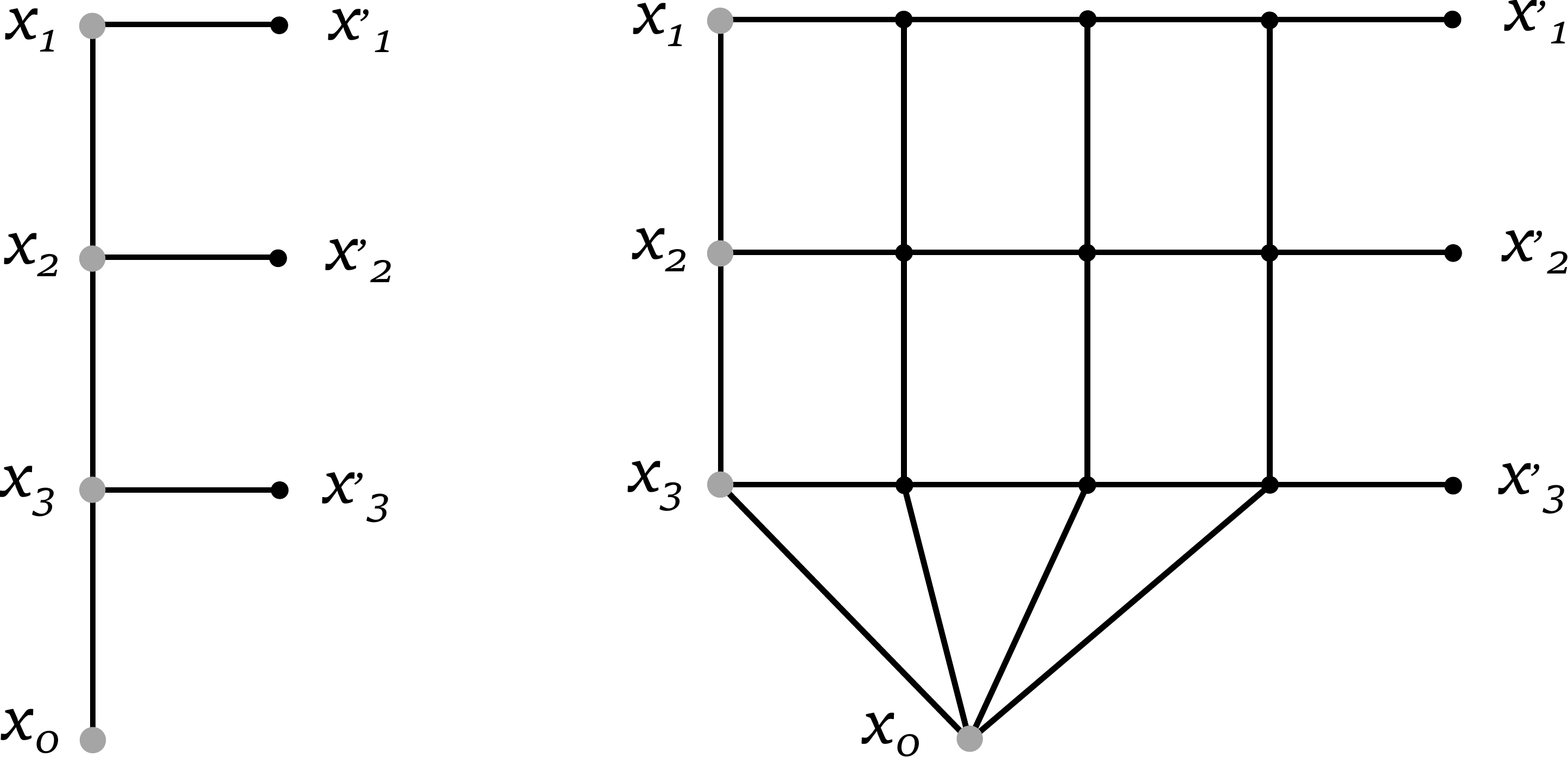}
  \caption{On the left the graph-building kernel $\mathbb{B}_3(\mathbf{x}|\mathbf{x}')$, where the lines are propagators $1/x_{ij}^2$, grey dots are external points and the black ones are integrated. On the right the portion of fishnet $(\mathbb{B}_3)^4$ with two fixed points $x_0$ (down) and $\infty$ (up).}
\label{gbuild}
\end{figure}
According to \eqref{tauN} the representation of the operator $(\mathbb{B}_N)^L $ over the separated variables factorizes completely a portion of size $N\times L$ of the planar fishnet diagram~\cite{Zamolodchikov:1980mb} in Fig.\ref{gbuild}, extending to a $4D$ space-time the analogue result in two-dimensions of \cite{Derkachov2018}. \\As a direct application of our results, we computed a specific set of four-point functions of Fishnet CFT \cite{Gurdogan:2015csr}, providing a direct check to formula (14) of \cite{Basso}, obtained via arguments of AdS/CFT correspondence \cite{Basso2015a,Fleury2017,Eden2017}.\\\\
In the next two sections we present the explicit construction of the eigenfunctions of the model \eqref{4Dlocal} by means of newly found integral identities.
 
\section{Generalized Star-triangle identity}
Our construction of a basis of eigenfunctions for $\mathbb{Q}_N(u)$ follows the logic outlined in \cite{Derkachov2014} for the two-dimensional model, and requires the formulation of certain conformal integral identities in $4D$.\\ First we consider a positive integer $M \leq N$ and set $x^{\mu}_0=0$ without loss of generality. We will denote $\mathbf{x}=(x_1,\dots,x_M)$, $\mathbf{x'}=(x'_{1},\dots,x'_{M-1})$. Let us introduce the tensors
\begin{align}
\label{Ctensor}
C^{\alpha\beta}_{\mu_1 \mu_{1'}\mu_{2} \dots \mu_{M}}=\langle \alpha |\boldsymbol{\bar\sigma}_{\mu_1} \boldsymbol{\sigma}_{\mu_1'} \boldsymbol{\bar \sigma}_{\mu_2}\cdots \boldsymbol{\bar \sigma}_{\mu_M}|\beta\rangle \,,
\end{align}
where the symbols $\boldsymbol{\sigma}$ and $\boldsymbol{\bar \sigma}$ are defined in terms of Pauli matrices
\begin{align}
\notag
\boldsymbol{\sigma}_0=\boldsymbol{\bar \sigma}_0 = \mathbbm{1},\,\,\boldsymbol{\sigma}_k=-\boldsymbol{\bar \sigma}_k = i\sigma_k,\,\,\,\,k=1,2,3\,.
\end{align}
The tensors \eqref{Ctensor} satisfy the light-cone condition
\begin{align}
\notag
t^{\mu_1\dots\mu_a}\,t^{\nu_1\dots\nu_a}\,C^{\alpha\beta}_{\mu_1\dots \mu_a \,\rho\dots \mu_{M}}C^{\alpha\beta}_{\nu_1 \dots\nu_a\,}{}^{\rho}{}_{\dots \nu_{M}} = 0\,,
\end{align}
where $t^{\mu_1\dots \mu_a}$ are auxiliary tensors and $a=1,1',\dots,M$.
This property allows to define a family of degree-$n$ homogeneous harmonic polynomials
\begin{align}
\label{Xtens}
C_{M}^{\alpha\beta}(\mathbf{x}|\mathbf{x'})^n= \langle\alpha|\mathbf{\bar x}_{11'}\mathbf{ {x}}_{1'2}\mathbf{\bar x}_{22'}\dots \mathbf{\bar x}_{M0}|\beta\rangle^n\,,
\end{align}
where $\mathbf{x}_{ij}=\boldsymbol{\sigma}_{\mu}x_{ij}^{\mu}/|x_{ij}|$ and  $\mathbf{\bar x}_{ij}=\boldsymbol{\bar \sigma}_{\mu} x_{ij}^{\mu}/|x_{ij}|$. Under a coordinate inversion $x^{\mu}\to x^{\mu}/x^2$ such harmonic polynomials transform covariantly and it follows that using \eqref{Xtens} it is possible to generalize the uniqueness - ``star-triangle" - relation for a conformal invariant vertex of three scalar propagators \cite{DEramo:1971hnd} (see also \cite{Isaev2003,Vasilev2004} and references therein) to any symmetric traceless representation. \\The core of the generalized identity is the mixing operator acting on a pair of symmetric spinors $|\alpha,\alpha'\rangle = |\alpha \rangle^{\otimes n} \otimes |\alpha'\rangle^{\otimes n'}$ of degrees $n$ and $n'$ as
\begin{align}
\label{Rmix}
&\notag\langle \alpha,\alpha' |\mathbf R_{n,n'}\left(z\right)|\beta,\beta'\rangle
 =  \frac{\Gamma(z+\frac{n-n'}{2})\Gamma(z+\frac{n'-n}{2})}
{\Gamma^2(z+\frac{n+n'}{2})}
\times\\&\times\partial_{s}^{n} \partial_{t}^{n'} (1+s \langle\alpha|\beta\rangle+t \langle \alpha'|\beta'\rangle +s t \langle \alpha|\beta'\rangle \langle \alpha' |\beta\rangle)^{z+\frac{n+n'}{2}}\, ,
\end{align}
where upon differentiation we set $s=t=0$. The operator defined by \eqref{Rmix} is a unitary solution of the Yang-Baxter equation and can be obtained via the fusion procedure \cite{Kulish1981} applied to the Yangian R-matrix $\mathbf{R}_{1,1}(z)$.\\
Under the uniqueness constraint $a+b+c=4$ and for any $n,n'\in \mathbb{N}$ the following identity holds
\begin{align}
\label{STR1}
\notag
\notag &\int d^4 x_4\, \frac{\langle \alpha | \mathbf{\bar x_{14}}\mathbf{x_{43}}| \beta\rangle^{n} \langle \alpha' |\mathbf{\bar x_{34}} \mathbf{x_{42}}| \beta'\rangle^{n'}}{(x_{14}^{2})^{a} \,(x_{24}^{2})^{b}\, (x_{34}^{2})^{c}}=\\&= \pi^2 \frac{(-1)^{n}\,A_{n,n'}(a,b,c)}{(x_{12}^2)^{(2-c)}(x_{13}^2)^{(2-b)}(x_{23}^2)^{(2-a)}}\times \notag \\& \times\frac{\langle \alpha \,\mathbf{\bar x_{12}}\mathbf{x_{23}},\alpha' |\mathbf R_{n,n'}\left(c-2\right)|\beta,\mathbf{\bar x_{31}} \mathbf{x_{12}}\,\beta'\rangle}{\left(c-1+\frac{n+n'}{2}\right)\left(2-c+\frac{n'-n}{2}\right)}\,.
\end{align}
with the coefficient $A_{n,n'}(a,b,c)=$
\begin{equation}
\notag
=\frac{\Gamma\left(2-a+\frac{n}{2}\right)\Gamma\left(2-b+\frac{n'}{2}\right)\Gamma\left(3-c+\frac{n'-n}{2}\right)}{\Gamma\left(a+\frac{n}{2}\right)\Gamma\left(b+\frac{n'}{2}\right)\Gamma\left(c-1+\frac{n'-n}{2}\right)}\,.
\end{equation}
Setting $n'=0$, the identity \eqref{STR1} is equivalent to (A.11) of \cite{Chicherin2013a}, and setting further $n=0$  it degenerates to the scalar identity \cite{DEramo:1971hnd}. \\
We point out that \eqref{STR1} is the four-dimensional versions of the $2D$ star-triangle relation which underlies the solution of the $SL(2,\mathbb{C})$ Heisenberg magnet as in \cite{Derkachov:2001yn,Derkachov2014}.
\section{Eigenfunctions construction}
The eigenfunctions of the open conformal chain \eqref{4Dlocal} can be obtained by a recursive procedure in the number of sites of the system.
First of all we introduce the integral operators $\hat \Lambda^{\alpha\beta}_{M,Y_a}=\langle \alpha |\hat\Lambda_{M,Y_a} |\beta\rangle$ \begin{align}
\label{lambda}
\hat \Lambda_{M,Y_a}^{\alpha\beta} \cdot\phi(\mathbf{x})=\int d\mathbf{x}' \,\Lambda^{\alpha\beta}_{M,Y_a}(\mathbf{x}|\mathbf{x'})\,\phi(\mathbf{x'})\,,
\end{align} through its kernel $\Lambda^{\alpha\beta}_{M,Y_a} (\mathbf{x}|\mathbf{x'})=\langle \alpha |\Lambda_{M,Y_a} (\mathbf{x}|\mathbf{x'})|\beta\rangle =$
\begin{align}
\notag
 = \frac{C_{M}^{\alpha
 \beta}(\mathbf{x}|\mathbf{x'})^{n_a}}{(x_{M0}^2)^{1+i\nu_a+i\lambda/2}}\prod_{a=1}^{M-1} \frac{(x_{a'a+1}^2)^{-1+i\nu_a+i\lambda/2}}{(x_{aa'}^2)^{1+i\nu_a-i\lambda/2}(x_{aa+1}^{2})^{i\lambda}}\,,
\end{align}
which at $M=1$ reduces to a conformal propagator of scaling dimension $\Delta=1+i\lambda/2+i\nu_a$ and tensor rank $n_a$
\begin{align}
\notag
\Lambda_{1,Y_a}^{\alpha\beta} (x_1) = \frac{\langle \alpha|\mathbf{\bar x}_{1}|{\beta}\rangle^{n_a}}{(x_1^2)^{1+i\nu_a+i\lambda/2}}\,.
\end{align}
Making use of \eqref{STR1} at $n=n_a,\,n'=0$ we verify that
\begin{align}
\label{exch1}
\mathbb{Q}_M(u) \,\hat\Lambda_{M,Y_a}^{\alpha\beta
}  = \tau_1(u,Y_a)\,\hat \Lambda_{M,Y_a}^{\alpha\beta} \,\mathbb{Q}_{M-1}(u)\,,
\end{align}
for any $M>1$, moreover 
\begin{align}
\label{M=1}
\mathbb{Q}_1(u)\, \Lambda_{1,Y_a}^{\alpha\beta}(x_1) =\tau_1(u,Y_a)\, \Lambda_{1,Y_a}^{\alpha\beta} (x_1)\,.
\end{align}
The iterative application of \eqref{exch1} for the length $M$ going from $N$ to $2$, together with the initial condition \eqref{M=1}, provides a recursive
definition of the eigenfunctions of the model with $N$ sites
\begin{align}
\label{eigenf}
\Psi^{\boldsymbol{\alpha\beta}} (\mathbf{Y}|\mathbf{x})= \hat \Lambda_{N,Y_N}^{\alpha_N\beta_N} \cdots\hat \Lambda_{2,Y_2\,}^{\alpha_2\beta_2}\cdot  \Lambda_{1,Y_1}^{\alpha_1\beta_1}\,\prod_{a=1}^N \frac{r(Y_a)^{a-1}}{\sqrt{2\pi^{2N+1}}}\,,
\end{align} 
where the last factor is a suitable normalization and
\begin{equation*}
r(Y)=\frac{\Gamma\left(Y-i\frac{\lambda}{2}\right)\Gamma\left(Y^*-i\frac{\lambda}{2}\right)}{\Gamma\left(Y+i\frac{\lambda}{2}\right)\Gamma\left(Y^*+i\frac{\lambda}{2}\right)}\,.
\end{equation*}
\begin{figure}
 \includegraphics[scale=0.33]{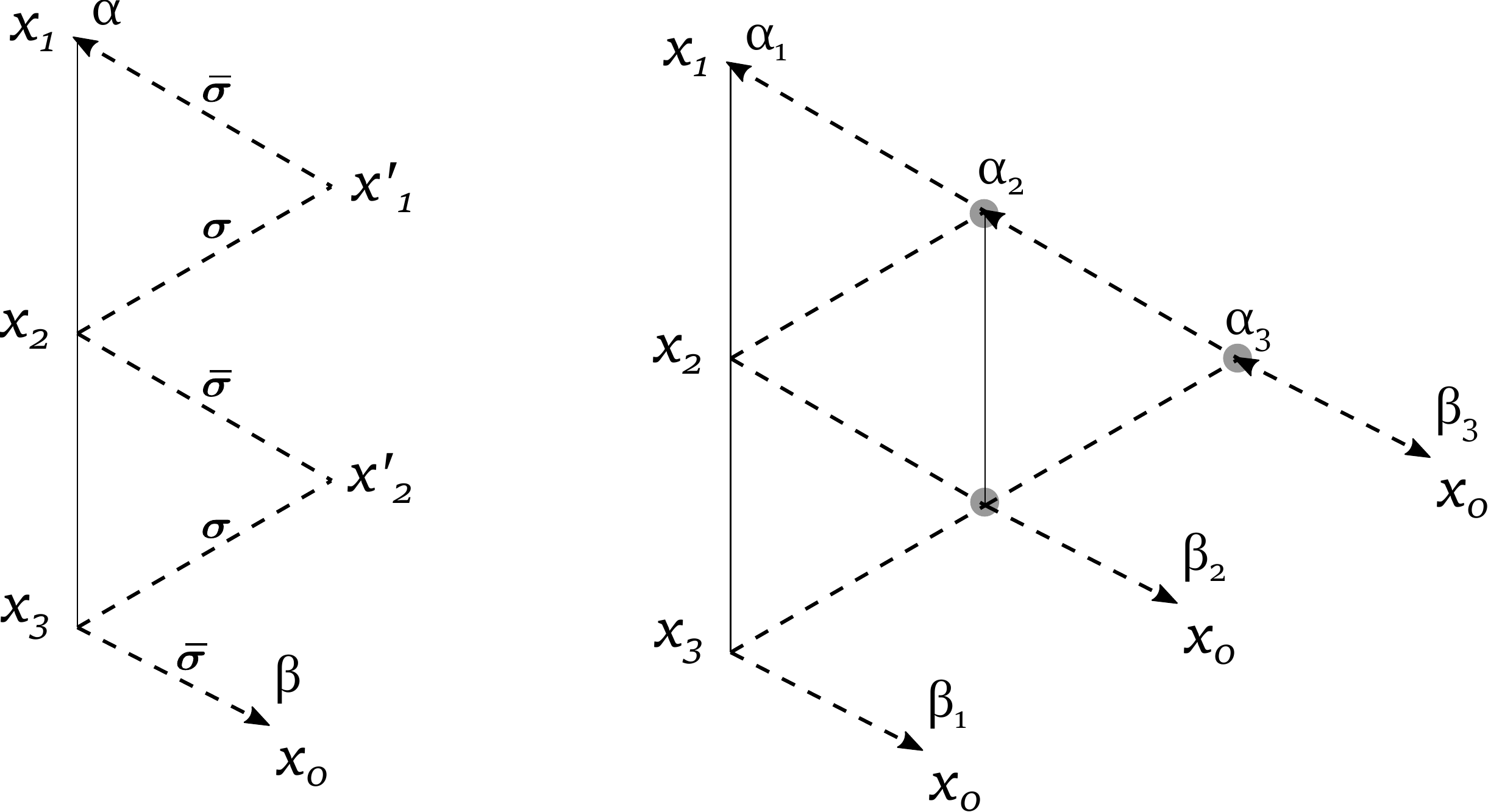}
 \label{eigen3}
  \caption{Graphic representation of the integral kernel $\Lambda^{\alpha,\beta}_{3,Y}(x_1,x_2,x_3|x_1',x_2')$ (left) and of the eigenfunction $\Psi^{\boldsymbol{\alpha\beta}}(\mathbf{Y}|x_1,x_2,x_3)$ (right). Solid lines denote $(x_{ij}^2)^{-i\lambda}$, while the dashed ones stand for the polynomials \eqref{Xtens} together with the denominators of type $(x_{i,i'}^2)$ and $(x_{i',i+1}^2)$ carrying the variables $\boldsymbol{\nu}$ in the power. The external arrows indicate symmetric spinors and the grey blobs are integrated points.}
\end{figure}
Such a function has a simple behavior in the permutation of two separated variables $(Y,\alpha,\beta)$, $(Y',\alpha',\beta')$, encoded by the exchange property
\begin{align}
\notag
\label{layerex}
&\hat\Lambda^{\alpha'\beta'}_{M,Y'}\cdot \hat\Lambda^{\alpha\beta}_{M-1,Y}=\langle \alpha', \alpha|\hat\Lambda_{M,Y'}\cdot \hat\Lambda_{M-1,Y}|\beta',\beta\rangle =\\&=\frac{r(Y)}{r(Y')}\,\langle \alpha,\alpha'|\mathbf{R}(z)^{\dagger}\,\hat\Lambda_{M,Y} \cdot \hat\Lambda_{M-1,Y'} \mathbf{R}(z)|\beta,\beta'\rangle\,,
\end{align}
where $z=i(\nu'-\nu)$ and $\mathbf{R}=\mathbf{R}_{n,n'}$. Any permutation of the separated variables in \eqref{eigenf} can be decomposed into elementary steps of type \eqref{layerex}, defining a representation of the symmetric group generators
\begin{equation*}
s_k \,\mathbf{Y}=(Y_1,\dots,Y_{k+1},Y_{k},\dots Y_N)\,,
\end{equation*}
on the space of symmetric spinors
\begin{align*}
&\mathbf{s}_k |\boldsymbol{\alpha}\rangle=\mathbf{R}_{n_k,n_{k+1}}(i\,\nu_{k+1,k})\,|\alpha_1,\dots,\alpha_{k+1},\alpha_{k},\dots,\alpha_N\rangle\,,
\end{align*}
and allowing to state the exchange symmetry
\begin{equation}
\label{Ysymm}
\Psi^{\boldsymbol{\alpha\beta}} (\mathbf{Y}|\mathbf{x})=\Psi^{ \boldsymbol{s_k(\alpha,\beta)}} (s_k \mathbf{Y}|\mathbf{x})\,.
\end{equation}
The scalar product of two eigenfunctions can be written according to \eqref{eigenf} in operatorial form, so that it can be reduced to $N$ factorized single-site contributions of the type
\begin{align*}
(\Lambda^{\alpha'\beta'}_{1,Y'})^{\dagger} \cdot  \Lambda^{\alpha\beta}_{1,Y}  = \frac{2\pi^3}{n+1}\,\delta_{n,n'}\delta(\nu-\nu') \langle\alpha|\alpha'\rangle^n\langle\beta|\beta'\rangle^n\,,
\end{align*}
by the iterative application of the property
\begin{align*}
&(\hat\Lambda^{\alpha'\beta'}_{M,Y'})^{\dagger}\cdot\hat\Lambda_{M,Y}^{\alpha\beta}= \langle \beta' ,\alpha|\hat \Lambda_{M,Y'}^{\dagger}\cdot\hat\Lambda_{M,Y}|\alpha',\beta\rangle =\frac{r(Y')}{r(Y)}\times \notag\\&  \times\pi^4 \frac{\text{Tr}_{n'}[\langle \alpha|\mathbf{R}(z)|\alpha'\rangle\,\hat\Lambda_{M-1,Y} \, \langle \beta'|\textbf{R}^{\dagger}(z)|\beta\rangle\,\hat\Lambda_{M-1,Y'}^{\dagger}]}{\left((\nu-\nu')^2+\frac{(n-n')^2}{4}\right)\left((\nu-\nu')^2+\frac{(n+n'+2)^2}{4}\right)}\,,
\end{align*}
valid under the assumption $Y\neq Y'$ and where the trace means the cyclic contraction of indices in the space of primed spinors. As result the scalar product of two functions \eqref{eigenf} takes the form of an orthogonality relation
\begin{align}
\label{orthog}
\frac{\mu(\mathbf{Y})^{-1}}{N!}\sum_{\pi \in \mathbb{S}_N} \delta(\mathbf Y-\pi(\mathbf{Y'})) \langle \boldsymbol{\alpha}|\boldsymbol \pi|\boldsymbol{\alpha'}\rangle \langle \boldsymbol{\beta'}|\boldsymbol\pi|\boldsymbol{\beta}\rangle\,,
\end{align}
where $\mathbb{S}_N$ are the permutations of $N$ objects and we introduced the compact notation
\begin{align*}
&\delta(\mathbf{Y}-\mathbf{Y}')= \prod_{a=1}^N \delta_{n_a,n_a'} \,\delta(\nu_a-\nu_a')\,.
\end{align*}
The relations \eqref{Ysymm},\eqref{orthog} allow to conjecture the completeness of the proposed eigenfunctions \eqref{eigenf} and to define the representation of separated variables as in \eqref{tilphi},\eqref{Ytrans}.

\section{Conformal Fishnet Integrals}
In analogy with the $2D$ results of~\cite{Derkachov2018}, employing the results of the previous sections we will compute exactly the four-point correlation function 
\begin{align}\label{G4}
G_{N,L}= \langle \text{Tr}[\phi_1^N(x_1)\phi_2^L(x_2)\phi_{1}^{\dagger N}(x_3)\phi_1^{\dagger L}(x_4)] \rangle \,,
\end{align}
for any $N$ and $L$, where $\phi_1(x)\,,\phi_2(x)$ are the two complex scalar $N_c\times N_c$ fields which appear in the Lagrangian of the conformal fishnet theory \cite{Gurdogan:2015csr} in four dimensions
\begin{align}
\notag
    {\cal L}_{\phi}&=  N_c\,\tr
    [\partial^{\mu}\phi_1^\dagger \,\partial_{\mu}\phi_1 + \partial^{\mu}\phi_2^\dagger \,\partial_{\mu}\phi_2 +(4 \pi)^2 \xi^2\phi_1^\dagger \phi_2^\dagger \phi_1\phi_2].
  \end{align}   In the planar limit \cite{tHooft:1973alw} $N_c\to \infty$ the only Feynmann diagram which contributes to the perturbative expansion in the coupling $\xi^2$ of $G_{N,L}$ is given by the integral
  \begin{equation}
\label{BD}
\medmuskip=-0.3mu
\thinmuskip=-0.3mu
\thickmuskip=-0.3mu
\int \frac{d\mathbf{z}}{(4\pi^2)^{NL}} \left(\prod_{a=0}^{N} \frac{1}{(z_{a,b}-z_{a+1,b})^2}\right)\left(\prod_{b=0}^{L}\frac{1}{(z_{a,b}-z_{a,b+1})^2}\right)\,,
 \end{equation}
where the integration measure is $d\mathbf z = \prod_{a,b=1}^{N,L} d^4 z_{a,b} $ and we set $z_{0b}=x_1,\,z_{N+1 b}=x_3,\,z_{a0}=x_4,\,z_{a L+1}=x_2$. Such a square-lattice integral can be expressed via the graph-building operator \eqref{graphB}. Indeed, starting from the fishnet diagram
\begin{equation}
 \label{BML}
 \medmuskip=0.8mu
\thinmuskip=0.8mu
\thickmuskip=0.8mu
F_{N,L}=\left(\prod_{a=1}^N z_{aa+1}^2\right)(\mathbb B_N)^{L+1}\left(\prod_{a=1}^N \delta^{(4)}(z'_a-z_a)\right)\,,
\end{equation}   
one can transform it to \eqref{BD} by the reductions of external points $z_a\to x_1$, $z'_a\to x_3$ followed by a conformal transformation. Therefore, as a functions  \(\mathcal G_{N,L}(u,v)\) of the cross-ratios \(u=x_{12}^2 x_{34}^2/(x_{13}^2 x_{24}^2)\)
and \(v=x_{14}^2x_{23}^2/(x_{13}^2 x_{24}^2)\), the planar limit of \eqref{G4} is equal to $F_{N,L}$ with reduced external points. According to \eqref{tauN} the integral kernel of $(\mathbb{B}_N)^L$ in the space of separated variables is factorized as
\begin{align}
\label{graphbY}
\widetilde{\mathbb{B}_{N}^L}(Y_1,\dots,Y_N)= \frac{1}{\pi^{2NL}}\prod_{a=1}^N\left[\frac{1}{4 \nu_a^2+(1+n_a)^2}\right]^L\,.
\end{align} In order to restore the $(u,v)$-dependence of \eqref{BD} one has first to expand the r.h.s. of \eqref{BML} over the eigenfunctions via the inverse transform \eqref{Ytrans}.
Then, by the appropriate reduction of the external points and upon integration of spinors and normalization by the bare correlator, we get
\begin{equation}
\notag
\mathcal G_{N,L}(u,v) = \sum_{\boldsymbol{n}\in \mathbb{Z}}\int {d\boldsymbol{\nu}~\mu(\mathbf Y)}\, \prod_{k=1}^{N} \frac{|x|^{-2i\nu_k} (\bar x/x )^{(n_k+1)/2}}{( \nu_k^2+(n_k+1)^2/4)^{L+N}}\,,
\end{equation}
where $u/v= x \bar x$, $v=1/\sqrt{(1-x)(1-\bar x)}$. After the redefinition $n_k\to a_k-1 $, $\nu_k \to u_k$, $x\to z$ it coincides with the result of \cite{Basso}.\\\\
\begin{figure}
 \includegraphics[scale=0.33]{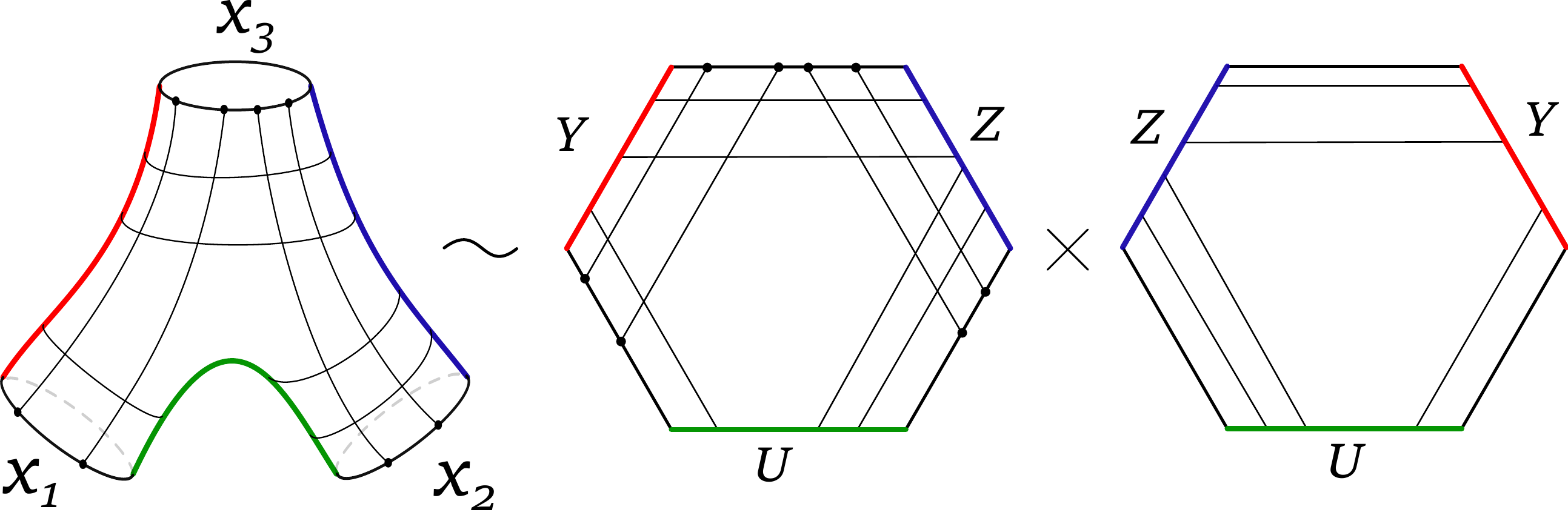}
  \caption{A Feynmann diagram contributing to the planar limit of $\langle\text{Tr}(\phi_1^2)(x_1) \text{Tr}(\phi_1^2)(x_2)\text{Tr}(\phi_1^{\dagger\, {4}})(x_3)\rangle$ at order $\xi^{28}$ and its decomposition into hexagons. Here $M_1=1$, $M_2=2$, $M_3=2$. Each color of a cut corresponds to the insertion of a different set of separated variables, as indicated on the hexagons.}
\label{3punc}
\end{figure}We shall conjecture further applications of the separated variables transform \eqref{Ytrans} to the computation of planar fishnet integrals. An interesting  example in this sense is provided by the three-point function of ``vacuum" operators \begin{align}
\label{3pt}
\langle\text{Tr}(\phi_1^N)(x_1) \text{Tr}(\phi_1^{L})(x_2)\text{Tr}(\phi_1^{\dagger\, {N+L}})(x_3)\rangle\,.
\end{align}
In the planar limit the perturbative expansion of \eqref{3pt} in the coupling constant consist of regular square lattice diagrams drawn on a three-punctured sphere $S^{2}$\textbackslash $\{x_1,x_2,x_3\}$ as explained in \cite{Basso2018} and exemplified in Fig.\ref{3punc}.  In the same spirit of ``hexagonalisation" techniques \cite{Eden2017,Fleury2017,Fleury2018,Basso2018} we perform three cuts on the diagram connecting the punctures, and insert along each cut a sum over the basis \eqref{eigenf}, labeled by the separated variables \begin{align}
\notag
(\mathbf{Y},\boldsymbol{\alpha},\boldsymbol{\beta})\,,\,\,\,(\mathbf{Z},\boldsymbol{\lambda},\boldsymbol{\chi})\,,\,\,\,(\mathbf{U},\boldsymbol{\kappa},\boldsymbol{\omega}) ,
\end{align}
where $Y_a=[\nu_a,n_a],\,Z_a=[\mu_a,m_a],\,U_a=[ \tau_a,t_a]$. Let $M_i$ be the number of $\phi_2\phi_2^{\dagger}$ wrappings around the puncture $x_i$ (see Fig.\ref{3punc}). The representation of the two hexagons over the separated variables reads
\begin{equation}
\notag
\medmuskip=-0.3mu
\thinmuskip=-0.2mu
\thickmuskip=-0.2mu
{|H|^2 \sim |\mathcal{A}|^2\,\,\,\prod_{a=1}^{M_1+M_3}\left[\frac{1}{\nu_a^2+\frac{(n_a+1)^2}{4}}\right]^{N}\,\ \prod_{b=1}^{M_2+M_3}\left[\frac{1}{\mu_b^2+\frac{(m_b+1)^2}{4}}\right]^{L}}\,,\end{equation}
and the form factor $\mathcal{A}$ is given by the overlapping of three eigenfunctions of type \eqref{eigenf} at different values of $x_0$
\begin{equation}
\label{formfac}
\medmuskip=1mu
\thinmuskip=1mu
\thickmuskip=1mu
\mathcal{A}=\int d \mathbf{z}\, d \mathbf{z'}\, d\mathbf{z''} \,\,\,\,\Psi^{\boldsymbol{\alpha}\boldsymbol{\beta}}_{\mathbf{Y}}(\mathbf{z},\mathbf{z}')\,\Psi_{\mathbf{Z}}^{\boldsymbol{\lambda}\boldsymbol{\chi}}(\mathbf{z},\mathbf{z''})\,\Psi_{\mathbf{U}}^{\boldsymbol{\kappa}\boldsymbol{\omega}}(\mathbf{z'},\mathbf{z''})\,,
\end{equation}
for $\mathbf{z}=(z_1,\dots,z_{M_3}),\,\mathbf{z'}=(z'_1,\dots,z'_{M_1}),$ and $\mathbf{z''}=(z''_1,\dots,z''_{M_2})$. Finally, the Feynmann integral is recovered by gluing the two hexagons via completeness sums
\begin{align*}
\sim \sum_{\mathbf{n},\mathbf{m},\mathbf{t}}\int  d \boldsymbol{\nu} \,d \boldsymbol{\mu}  \,d \boldsymbol{\tau}\, {\mu}(\mathbf{Y})\, {\mu}(\mathbf{Z}) \,{\mu}(\mathbf{U})\int D\boldsymbol{\alpha}\cdots D\boldsymbol{\omega} \,|H|^2\,.
\end{align*}
An interesting reduction of the correlator \eqref{3pt} is obtained setting $L=0$ and degenerating it to the two-point function $\langle\text{Tr}(\phi_1^N)(x_1) \text{Tr}(\phi_1^{\dagger\, {N}})(x_3)\rangle $, for which the planar fishnet lies on a cylinder and it is conformally equivalent to a ``wheel" diagram \cite{Broadhurst:1985vq,Panzer:2015ida,Gurdogan:2015csr,Gromov:2017cja}. \\As a general fact the diagrams describing the planar limit of \eqref{3pt} develop UV divergences, which in our representation should be contained in the form factor \eqref{formfac}. The elaboration of a regularization technique at this level is an intriguing task as it would enable the direct computation of several conformal data in the Fishnet CFT at finite order in the coupling.
\section{Conclusions}
We formulated and solved the spin chain of $SO(1,5)$ conformal spins for any number of sites $N$ and for open boundary conditions, in the principal series representation of zero spin~\cite{Dobrev1977}. Its integrability is realized by a commuting family of spectral parameter-dependent operators $\mathbb{Q}_N(u)$ which generate the conserved charges of the model. The spectrum of the model is separated into $N$ symmetric contributions, each depending on quantum numbers which for this reason we call separated variables. We explained how to construct the eigenfunctions and prove their orthogonality, extending the logic of \cite{Derkachov2014} to a four dimensional space-time by means of new integral indentities which generalize the star-triangle relation \cite{DEramo:1971hnd} to symmetric traceless tensors. \\Our results can be analytically continued from the representation of the principal series to real scaling dimensions, recovering the graph-building operator - introduced in $2D$ by the authors and V.~Kazakov \cite{Derkachov2018} - for the Feynmann diagrams of Fishnet CFT \cite{Gurdogan:2015csr,Caetano:2016ydc}.
The variant of this graph-builder with periodic boundary was first introduced in \cite{Gurdogan:2015csr} and coincides with the $\hat{B}$-operator of the Fishchain holographic model \cite{Gromov2019c,Gromov2019b,Gromov2019a}. Following the same steps as \cite{Derkachov2018}, we computed the planar limit of the fishnet correlator studied by B.~Basso and L.~Dixon providing a direct check of the formula (14) of \cite{Basso}.\\ The separation of variables (SoV) for non-compact spin magnets is a topic which recently attracted great attention~\cite{Maillet2018,Maillet2019a,Gromov2017,Cavaglia2019,Ryan2019,Gromov2019}, and SoV features appear in remarkable results of AdS/CFT integrability, for instance~\cite{Kostov2019a,Jiang2019a}. It has not escaped our notice that the properties of the proposed eigenfunctions immediately suggest their role in the SoV of the periodic $SO(1,5)$ spin chain~\cite{Chicherin2013a}, in full analogy with~\cite{Derkachov:2001yn}. Moreover it would be interesting to apply our methods to the computation of other classes of Feynmann integrals, for example introducing fermions as in \cite{Kazakov2019,Pittelli2019}, or considering any space-time dimension and extending our results to the theory proposed in \cite{Kazakov}. In the latter context, the functions \eqref{eigenf} for $N=2$ sites have been derived in a somewhat different form and applied to the formulation of the Thermodynamic Bethe Ansatz equations~\cite{Basso2019a}.\\
Finally we have conjectured how, by means of a cutting-and-gluing procedure inspired by \cite{Basso2018}, certain planar two- and three-point functions of the Fishnet CFT at finite coupling get factorized into simple contributions over the separated variables. This observation puts as a compelling future task the regularization of such formulas, in order to compare the results based on the AdS/CFT correspondence to a direct computation.
\begin{acknowledgments}
\section*{Acknowledgments}
\label{sec:acknowledgments}

We thank B.~Basso, A.~Manashov, for useful discussions and G.~Ferrando and D-l.~Zhong for comments on the manuscript. We are grateful to V.~Kazakov for participating in the initial stages of the project. The work of S.D. was supported
by the RFBR grant no. 17-01-00283a. The work of E. O. was funded by the German Science Foundation (DFG) under the Research Training Group 1670 and under Germany's Excellence Strategy -- EXC 2121 ``Quantum Universe" -- 390833306.

\end{acknowledgments}

\bibliographystyle{apsrev4-1}
\bibliography{SoV_letter}

\end{document}